\documentstyle[12pt]{article}

  \def\pp{{\mathchoice
          {
              \kern 1pt%
              \raise 1pt
              \vbox{\hrule width5pt height0.4pt depth0pt
                    \kern -2pt
                    \hbox{\kern 2.3pt
                          \vrule width0.4pt height6pt depth0pt
                          }
                    \kern -2pt
                    \hrule width5pt height0.4pt depth0pt}%
                    \kern 1pt
           }
            {
              \kern 1pt%
              \raise 1pt
              \vbox{\hrule width4.3pt height0.4pt depth0pt
                    \kern -1.8pt
                    \hbox{\kern 1.95pt
                          \vrule width0.4pt height5.4pt depth0pt
                          }
                    \kern -1.8pt
                    \hrule width4.3pt height0.4pt depth0pt}%
                    \kern 1pt
            }
            {
              \kern 0.5pt%
              \raise 1pt
              \vbox{\hrule width4.0pt height0.3pt depth0pt
                    \kern -1.9pt  
                    \hbox{\kern 1.85pt
                          \vrule width0.3pt height5.7pt depth0pt
                          }
                    \kern -1.9pt
                    \hrule width4.0pt height0.3pt depth0pt}%
                    \kern 0.5pt
            }
            {
              \kern 0.5pt%
              \raise 1pt
              \vbox{\hrule width3.6pt height0.3pt depth0pt
                    \kern -1.5pt
                    \hbox{\kern 1.65pt
                          \vrule width0.3pt height4.5pt depth0pt
                          }
                    \kern -1.5pt
                    \hrule width3.6pt height0.3pt depth0pt}%
                    \kern 0.5pt
            }
        }}

  \def\mm{{\mathchoice
                       {
                             \kern 1pt
               \raise 1pt    \vbox{\hrule width5pt height0.4pt depth0pt
                                  \kern 2pt
                                  \hrule width5pt height0.4pt depth0pt}
                             \kern 1pt}
                       {
                            \kern 1pt
               \raise 1pt \vbox{\hrule width4.3pt height0.4pt depth0pt
                                  \kern 1.8pt
                                  \hrule width4.3pt height0.4pt depth0pt}
                             \kern 1pt}
                       {
                            \kern 0.5pt
               \raise 1pt
                            \vbox{\hrule width4.0pt height0.3pt depth0pt
                                  \kern 1.9pt
                                  \hrule width4.0pt height0.3pt depth0pt}
                            \kern 1pt}
                       {
                           \kern 0.5pt
             \raise 1pt  \vbox{\hrule width3.6pt height0.3pt depth0pt
                                  \kern 1.5pt
                                  \hrule width3.6pt height0.3pt depth0pt}
                           \kern 0.5pt}
                       }}

\catcode`@=11
\def\un#1{\relax\ifmmode\@@underline#1\else
        $\@@underline{\hbox{#1}}$\relax\fi}
\catcode`@=12

\let\du=\du                     

\def\a{\alpha}
\def\b{\beta}

\def\d{\delta}

\def\f{\phi}

\def\h{\eta}

\def\j{\psi}

\def\l{\lambda}
\def\m{\mu}
\def\n{\nu}

\def\p{\pi}
\def\q{\theta}
\def\r{\rho}
\def\s{\sigma}
\def\t{\tau}

\def\F{\Phi}

\def\L{\Lambda}

\def\ve{\varepsilon}

\def\cy{{\cal Y}}

\def\bo{{\raise-.3ex\hbox{\large$\Box$}}}               
\def\pa{\partial}                                       
\def\pr{\prod}                                          
\def\TH{{\raise.2ex\hbox{$\displaystyle \bigodot$}\mskip-4.7mu \llap H \;}}
\def\face{{\raise.2ex\hbox{$\displaystyle \bigodot$}\mskip-2.2mu \llap {$\ddot
        \smile$}}}                                      

\def\sp#1{{}^{#1}}                              
\def\Tilde#1{\widetilde{#1}}                    
\def\Bar#1{\overline{#1}}                       
\def\leftrightarrowfill{$\mathsurround=0pt \mathord\leftarrow \mkern-6mu
        \cleaders\hbox{$\mkern-2mu \mathord- \mkern-2mu$}\hfill
        \mkern-6mu \mathord\rightarrow$}
\def\dvec#1{\vbox{\ialign{##\crcr
        \leftrightarrowfill\crcr\noalign{\kern-1pt\nointerlineskip}
        $\hfil\displaystyle{#1}\hfil$\crcr}}}           
\def\dt#1{{\buildrel {\hbox{\LARGE .}} \over {#1}}}     

\def\frac#1#2{{\textstyle{#1\over\vphantom2\smash{\raise.20ex
        \hbox{$\scriptstyle{#2}$}}}}}                   
\def\sfrac#1#2{{\vphantom1\smash{\lower.5ex\hbox{\small$#1$}}\over
        \vphantom1\smash{\raise.4ex\hbox{\small$#2$}}}} 
\def\bfrac#1#2{{\vphantom1\smash{\lower.5ex\hbox{$#1$}}\over
        \vphantom1\smash{\raise.3ex\hbox{$#2$}}}}       
\def\afrac#1#2{{\vphantom1\smash{\lower.5ex\hbox{$#1$}}\over#2}}    

\def\[{\lfloor{\hskip 0.35pt}\!\!\!\lceil}
\def\]{\rfloor{\hskip 0.35pt}\!\!\!\rceil}

\def\du#1#2{_{#1}{}^{#2}}
\def\ud#1#2{^{#1}{}_{#2}}

\def\ha{{\fracmm12}}
\def\tr{{\rm tr}}

\def\un{\underline}
\def\fracmm#1#2{{{#1}\over{#2}}}

\def\low#1{{\raise -3pt\hbox{${\hskip 0.75pt}\!_{#1}$}}}

\def\Dot#1{\buildrel{_{_{\hskip 0.01in}\bullet}}\over{#1}}
\def\dt#1{\Dot{#1}}

\def\Tilde#1{{\widetilde{#1}}\hskip 0.015in}

\newskip\humongous \humongous=0pt plus 1000pt minus 1000pt
\def\caja{\mathsurround=0pt}
\def\eqalign#1{\,\vcenter{\openup2\jot \caja
        \ialign{\strut \hfil$\displaystyle{##}$&$
        \displaystyle{{}##}$\hfil\crcr#1\crcr}}\,}
\newif\ifdtup

\def\ref#1{$\sp{#1)}$}

\def\pl#1#2#3{Phys.~Lett.~{\bf {#1}B} (19{#2}) #3}
\def\np#1#2#3{Nucl.~Phys.~{\bf B{#1}} (19{#2}) #3}

\def\pr#1#2#3{Phys.~Rev.~{\bf D{#1}} (19{#2}) #3}

\def\mpl#1#2#3{Mod.~Phys.~Lett.~{\bf A{#1}} (19{#2}) #3}

\parskip=\medskipamount                 
\thicklines                         

\begin{document}


\thispagestyle{empty}               

\def\border{                                            
        \setlength{\unitlength}{1mm}
        \newcount\xco
        \newcount\yco
        \xco=-24
        \yco=12
        \begin{picture}(140,0)
        \put(-20,11){\tiny Institut f\"ur Theoretische Physik Universit\"at
Hannover~~ Institut f\"ur Theoretische Physik Universit\"at Hannover~~
Institut f\"ur Theoretische Physik Hannover}
        \put(-20,-241.5){\tiny Institut f\"ur Theoretische Physik Universit\"at
Hannover~~ Institut f\"ur Theoretische Physik Universit\"at Hannover~~
Institut f\"ur Theoretische Physik Hannover}
        \end{picture}
        \par\vskip-8mm}

\def\headpic{                                           
        \indent
        \setlength{\unitlength}{.8mm}
        \thinlines
        \par
        \begin{picture}(29,16)
        \put(75,16){\line(1,0){4}}
        \put(80,16){\line(1,0){4}}
        \put(85,16){\line(1,0){4}}
        \put(92,16){\line(1,0){4}}

        \put(85,0){\line(1,0){4}}
        \put(89,8){\line(1,0){3}}
        \put(92,0){\line(1,0){4}}

        \put(85,0){\line(0,1){16}}
        \put(96,0){\line(0,1){16}}
        \put(92,16){\line(1,0){4}}

        \put(85,0){\line(1,0){4}}
        \put(89,8){\line(1,0){3}}
        \put(92,0){\line(1,0){4}}

        \put(85,0){\line(0,1){16}}
        \put(96,0){\line(0,1){16}}
        \put(79,0){\line(0,1){16}}
        \put(80,0){\line(0,1){16}}
        \put(89,0){\line(0,1){16}}
        \put(92,0){\line(0,1){16}}
        \put(79,16){\oval(8,32)[bl]}
        \put(80,16){\oval(8,32)[br]}

        \end{picture}
        \par\vskip-6.5mm
        \thicklines}

\border\headpic {\hbox to\hsize{
\vbox{\noindent DESY 98 -- 138 \hfill September 1998 \\
ITP--UH--22/98 \hfill hep-th/9809121 }}}

\noindent
\vskip1.3cm

\begin{center}
{\large\bf A manifestly N=2 supersymmetric Born-Infeld action}
\footnote{Supported in part by the `Deutsche Forschungsgemeinschaft'}\\
\vglue.3in

Sergei V. Ketov \footnote{
On leave from:
High Current Electronics Institute of the Russian Academy of Sciences, 
Siberian 
\newline ${~~~~~}$ Branch, Akademichesky~4, Tomsk 634055, Russia}

{\it Institut f\"ur Theoretische Physik, Universit\"at Hannover}\\
{\it Appelstra\ss{}e 2, 30167 Hannover, Germany}\\
{\sl ketov@itp.uni-hannover.de}
\end{center}
\vglue.2in
\begin{center}
{\Large\bf Abstract}
\end{center}

A manifestly N=2 supersymmetric completion of the four-dimensional 
Nambu-Goto-Born-Infeld action, which is self-dual with respect to 
electric-magnetic duality, is constructed in terms of an abelian N=2 
superfield strength $W$ in the conventional N=2 superspace. A relation 
to the known N=1 supersymmetric Born-Infeld action in N=1 superspace is 
established. The action found can be considered either as the Goldstone 
action associated with a partial breaking of N=4 supersymmetry down to N=2, 
with the N=2 vector superfield being a Goldstone field, or, equivalently, 
as the gauge-fixed superfield action of a D-3-brane in flat six-dimensional 
ambient spacetime.

\newpage

The {\it Born-Infeld} (BI) action in the four-dimensional spacetime with a 
metric $g_{\m\n}$,
$$ S_{\rm BI}=\fracmm{1}{b^2}\int d^4x\,\left\{\sqrt{-g}-\sqrt{-\det(g_{\m\n}
+bF_{\m\n})}\,\right\}~,\eqno(1)$$
was originally suggested \cite{bi} as a non-linear version of Maxwell theory 
with the abelian vector gauge field strength 
$F_{\m\n}=\pa_{\m}A_{\n}-\pa_{\n}A_{\m}$. The BI action also arises
as the bosonic part of the open superstring effective action \cite{open}, 
an amalgam of the Born-Infeld and Nambu-Goto (NG) actions is just the  
gauged-fixed D-3-brane effective action \cite{lei,aps}. In string theory
$b=2\p\a'$, whereas we take $b=1$ is our paper for notational simplicity. 
In a flat spacetime the BI lagrangian takes the form
$$ L_{\rm BI}=1-\sqrt{-\det(\h_{\m\n}+F_{\m\n})}
=-\frac{1}{4}F^2 + \frac{1}{32}\left[
(F^2)^2 +(F\Tilde{F})^2\right] +O(F^6)~,\eqno(2)$$
where $\h_{\m\n}$ is Minkowski metric, and we have used the notation
$$ \Tilde{F}_{\m\n}=\frac{1}{2}\ve\du{\m\n}{\r\l}F_{\r\l}~,
\quad F^2=F_{\m\n}F^{\m\n}~,
\quad{\rm and}\quad F\Tilde{F}=F^{\m\n}\Tilde{F}_{\m\n}~.\eqno(3)$$
The leading terms explicitly written down on the r.h.s. of eq.~(2) agree with 
the {\it Euler-Heisenberg} (EH) effective lagrangian in supersymmetric scalar 
QED with the parameter $b^2=e^4/(24\p^2 m^4)$ \cite{eh}. 

For a later use, let's also introduce complex combinations of $F_{\m\n}$ and 
its dual $\Tilde{F}_{\m\n}$,
$$ F^{\pm}=\frac{1}{2}(F \pm i\Tilde{F})~,\eqno(4)$$
which satisfy the identities
$$ (F^{\pm})^2=\frac{1}{2}(F^2 \pm iF\Tilde{F}) \quad{\rm and}\quad
4(F^+)^2(F^-)^2=(F^2)^2 +(F\Tilde{F})^2~.\eqno(5)$$

A lagrangian {\it magnetically} dual to the BI one is obtained via the 
first-order action
$$ L_1=L_{\rm BI}(F) + 
\frac{1}{2}\Tilde{A}_{\m}\ve^{\m\n\l\r}\pa_{\n}F_{\l\r}~,\eqno(6)$$
where $\Tilde{A}_{\m}$ is a (dual) magnetic vector potential. $\Tilde{A}_{\m}$
enters eq.~(6) as the Lagrange multiplier and thus enforces the Bianchi 
identity $\ve^{\m\n\l\r}\pa_{\n}F_{\l\r}=0$. Varying eq.~(6) with respect to 
$F_{\m\n}$ instead, solving the resulting algebraic equation for $F_{\m\n}$ as
a function of $\Tilde{A}_{\m}$ and substituting a solution back into eq.~(6) 
yields a magnetically dual BI action which has {\it the same} form as in 
eq.~(1), but in terms of the magnetically dual field strength 
${}^*F_{\m\n}=\pa_{\m}\Tilde{A}_{\n}-\pa_{\n}\Tilde{A}_{\m}$  \cite{schr,gr}. 
The easiest way to check it is to take advantage of the Lorentz invariance of 
the BI action (1) by putting $F_{\m\n}$ into the form
$$F_{\m\n}=\left(\begin{array}{cccc} 0 & f_1 & 0 & 0 \cr
-f_1 & 0 & 0 & 0 \cr 0 & 0 & 0 & f_2 \cr 0 & 0 & -f_2 & 0 \cr
\end{array}\right)\eqno(7)$$
in terms of its eigenvalues $(f_1,f_2)$, and similarly for 
$\Tilde{{}^*F_{\m\n}}=\frac{1}{2}\ve\du{\m\n}{\l\r}\pa_{\l}\Tilde{A}_{\r}$ in 
terms of its eigenvalues $(\l_1,\l_2)$. It is now straightforward to
verify that, given the first-order lagrangian $L=\sqrt{(1-f^2_1)(1+f^2_2)} +
\l_1f_1+\l_2f_2$ its magnetically dual one is given by  
${}^*L=\sqrt{(1-\l^2_1)(1+\l^2_2)}$ indeed \cite{tse}. The non-gaussian BI 
lagrangian (1) is uniquely fixed by the electric-magnetic self-duality 
requirement alone, when one also insists on the correct (Maxwell) weak field 
theory limit. In general, there exists a family of self-dual lagrangians 
parameterized by one variable, all being solutions to a first-order 
Hamilton-Jacobi partial differential equation \cite{gr}.

One expects that 4d supersymmetry should be also compatible with the 
electric-magnetic self-duality of the BI action (1) since the 
electric-magnetic diality transformations commute with spacetime symmetries. 
It turns out to be the case indeed, after promoting the abelian vector field 
$A_{\m}$ to a vector supermultiplet and supersymmetrizing both the BI action 
and its duality transformations (see below). As a preparation for a 
supersymmetrization of eq.~(1), let' use the identity 
$$ -\det(\h_{\m\n}+F_{\m\n})=
1 +\frac{1}{2}F^2 +\frac{1}{8}(F^2)^2-\frac{1}{4}F^4
=1 +\frac{1}{2}F^2 -\frac{1}{16}(F\Tilde{F})^2
~,\eqno(8)$$
where we have introduced $F^4\equiv F_{\m\n}F^{\n\l}F_{\l\r}F^{\r\m}$, as well
as yet another identity
$$ F^4= \frac{1}{4}(F\Tilde{F})^2 + \frac{1}{2}(F^2)^2~.\eqno(9)$$
The BI action in any dimension $d$ is often put into another form \cite{tse},
$$ L_4=\int d^dx\,\left[ 1+\fracmm{1}{2}\L \det(\h_{\m\n}+F_{\m\n})
-\fracmm{1}{2\L}\right]~,\eqno(10)$$
where the auxiliary field $\L$ has been introduced. However, eq.~(10) does not
seem to be a good starting point for supersymmetrization in $d=4$ spacetime 
dimensions. Instead, let's define 
$$ A\equiv -\frac{1}{4}F^2 \quad {\rm and}\quad 
B\equiv \frac{1}{32}\left[ (F^2)^2+(F\Tilde{F})^2\right]~,\eqno(11)$$
which essentially represent the Maxwell limit of the BI action and the square 
of the Maxwell energy-momentum tensor, respectively, and rewrite eqs.~(8) and 
(2), respectively, as
$$ -\det(\h_{\m\n}+F_{\m\n})=(1-A)^2-2B~,\eqno(12)$$
and
$$ 
L_{\rm BI}=1-\sqrt{-\det(\h_{\m\n}+F_{\m\n})}
=A+B+\ldots \equiv A +B Y(A,B)~,\eqno(13)$$
where the function $Y(A,B)$ has been introduced as a solution to the 
quadratic equation
$$ By^2+2(A-1)y+2=0~.\eqno(14)$$
It is not difficult to check that
$$ Y(A,B)=\fracmm{1-A-\sqrt{(1-A)^2-2B}}{B}~. \eqno(15)$$

A manifestly N=1 supersymmetric completion of the bosonic four-dimensional BI 
theory (2) is known \cite{cf,bg}, whereas we want to find an extended version 
of the NGBI action with linearly realized N=2 supersymmetry. The corresponding
action can be most easily found in the conventional N=2 superspace 
parameterized by the coordinates 
$Z^M=(x^{\m},\q^{\a}_i,\bar{\q}^{\dt{\a}i})$, where $\m=0,1,2,3$, $\a=1,2$, 
$i=1,2$, and $\Bar{\q^{\a}_i}=\bar{\q}^{\dt{\a}i}$, in terms of a restricted 
chiral N=2 superfield $W$ representing the N=2 supersymmetric abelian gauge 
field strength \cite{wess,oldrev}. The N=2 superspace approach automatically 
implies manifest (i.e. linearly realized) N=2 extended supersymmetry. Our 
approach cannot, however, be used to construct a NGBI action with manifest N=4 
supersymmetry since a 4d gauge theory with linearly realized N=4 supersymmetry
merely exists in its on-shell version (in N=4 superspace).

The restricted N=2 chiral superfield $W$ is an off-shell irreducible N=2 
superfield satisfying the N=2 superspace constraints
$$ \bar{D}_{\dt{\a}i}W=0~,\qquad D^4W=\bo \bar{W}~,\eqno(16)$$
where we have used the following realization of the supercovariant N=2 
superspace derivatives (with vanishing central charge) \cite{oldrev}:
$$ D^i_{\a}=\fracmm{\pa}{\pa\q^{\a}_i} +i\bar{\q}^{\dt{\a}i}\pa_{\a\dt{\a}}~,
\quad
\bar{D}_{\dt{\a}i}= -\fracmm{\pa}{\pa\bar{\q}^{\dt{\a}i}} 
-i\q^{\a}_i\pa_{\a\dt{\a}}~;\quad D^4 
\equiv \frac{1}{12}D^{i\a}D^j_{\a}D^{\b}_iD_{j\b}~.\eqno(17)$$
The first constraint of eq.~(16) is just the N=2 generalization of the usual 
N=1 chirality condition, whereas the second one can be considered as the 
generalized {\it reality} condition \cite{sg,oldrev} which has no analogue in 
N=1 superspace. A solution to eq.~(16) in N=2 chiral superspace
$(y^{\m}=x^{\m}-\frac{i}{2}\q_i^{\a}\s^{\m}_{\a\dt{\a}}\bar{\q}^{\dt{\a}i},
\q_{\b}^j)$ reads
$$\eqalign{
W(y,\q)~=~& a(y) + \q^{\a}_i\j^i_{\a}(y)-\ha\q^{\a}_i\vec{\t}\ud{i}{j}
\q^j_{\a}\cdot\vec{D}(y)\cr
~&  +\frac{i}{8}\q_i^{\a}(\s^{\m\n})\du{\a}{\b}\q_{\b}^iF_{\m\n}(y)
-i(\q^3)^{i\a}\pa_{\a\dt{\b}}\bar{\j}_i^{\dt{b}}(y)+\q^4\bo \bar{a}(y)~,\cr}
\eqno(18)$$
where we have introduced a complex scalar $a$, a chiral spinor doublet $\j$, 
a real isovector
$\vec{D}=\ha(\vec{\t})\ud{i}{j}D\ud{j}{i}\equiv \ha\tr(\vec{\t}D)$, 
$\tr(\t_m\t_n)=2\d_{mn}$,
and a real antisymmetric tensor $F_{\m\n}$ as the field components of $W$, 
while $F_{\m\n}$ has to satisfy `Bianchi identity' \cite{oldrev}
$$ \ve^{\m\n\l\r}\pa_{\n}F_{\l\r}=0 ~,\eqno(19)$$
whose solution is just given by $F_{\m\n}=\pa_{\m}A_{\n}-\pa_{\n}A_{\m}$ in 
terms of a vector gauge field $A_{\m}$ subject to the gauge transformations 
$\d A_{\m}=\pa_{\m}\l$. The N=2 supersymmetry transformation laws for the 
components can be found e.g., in ref.~\cite{oldrev}.

The choice of variables made in eq.~(11) was actually dictated by the 
observation that their N=2 supersymmetric extensions are easy to construct. 
The well-known N=2 supersymmetric extension of the
Maxwell lagrangian $A=-\frac{1}{4}F^2_{\m\n}$ is given by
$$ -\frac{1}{2} \int d^4\q\,W^2=
-a\bo\bar{a}-\frac{i}{2}\j^{\a}_j\pa_{\a\dt{\a}}\bar{\j}^{\dt{\a}j}
-\frac{1}{2}(F^+)^2 +\frac{1}{2}\vec{D}^2~.\eqno(20)$$
The Maxwell energy-momentum tensor squared $(B)$ is also easily extended in 
N=2 superspace to the N=2 supersymmetric EH lagrangian
$$ \int d^4\q d^4\bar{\q}\,W^2\bar{W}^2=(F^+)^2(F^-)^2 
+(\vec{D}^2)^2 -\vec{D}^2F^2+\ldots~.\eqno(21)$$
This N=2 supersymmetric generalization of the EH lagrangian also arises as
the leading (one-loop) non-holomorphic (non-BPS) contribution to the N=2 gauge 
low-energy effective action of a charged hypermultipet minimally interacting 
with an N=2 Maxwell multiplet \cite{buch,krev}.  

The gauge-invariant N=2 superfield strength squared, $W^2$, is an 
N=2 chiral but not a restricted N=2 chiral superfield. As is clear from 
eq.~(20), the first component of an N=2 anti-chiral superfield  
$K\equiv D^4W^2$ takes the form
$$ K \equiv D^4W^2= 2a\bo\bar{a}+(F^+)^2-\vec{D}^2+\ldots~. \eqno(22)$$

It is now straightforward to N=2 supersymmetrize the BI lagrangian (13) by 
engineering an N=2 superspace invariant ({\it cf.} ref.~\cite{cf})
$$ L = -\frac{1}{2} \int d^4\q\, W^2 -\frac{1}{8}
\int d^4\q d^4\bar{\q}\,\cy(K,\bar{K}) W^2\bar{W}^2~,
\eqno(23)$$
whose `formfactor' $\cy(K,\bar{K})$ is dictated by the structure function 
(15), i.e.
$$ \cy(K,\bar{K})=Y(A,B)~.\eqno(24)$$
Note that the vector-dependent contributions to the first components of $K$ 
and $\bar{K}$ are simply related to $A$ und $B$ as
$$ K+\bar{K}=-4A~,\qquad K\bar{K}=B~,\eqno(25)$$
i.e. they are just the roots of yet another quadratic equation
$$ k^2+4Ak+B=0~.\eqno(26)$$
We find
$$\eqalign{
 \cy(K,\bar{K})~=~ &
\fracmm{1+\frac{1}{4}(K+\bar{K})-\sqrt{(1+\frac{1}{4}K+\frac{1}{4}\bar{K})^2
-2K\bar{K}}}{K\bar{K}} \cr
~=~ & 1-\frac{1}{4}(K+\bar{K}) +O(K^2)~.\cr} \eqno(27)$$

The proposed action
$$\eqalign{
S[W,\bar{W}] ~& = -\frac{1}{2} \int d^4x d^4\q\, W^2 -\frac{1}{8}
\int d^4x d^4\q d^4\bar{\q}\,\cy(K,\bar{K}) W^2\bar{W}^2\cr
~& = -\frac{1}{2} \int d^4x d^4\q \left\{  W^2 +\frac{1}{4}\bar{D}^4\left[
\cy(K,\bar{K}) W^2\bar{W}^2\right]\right\} \cr
~& = -\frac{1}{2} \int d^4x d^4\q \, W^2_{\rm improved}~,\cr}\eqno(28)$$
can be nicely rewritten to the `non-linear sigma-model' form
$$ S[W,\bar{W}] = - \frac{1}{2} \int d^4x d^4\q\, X ~,\eqno(29)$$
where the N=2 chiral superfield $X\equiv  W^2_{\rm improved}$ 
has been introduced as a solution to the non-linear N=2 superfield constraint
$$ X = \frac{1}{4}X\bar{D}^4\bar{X} + W^2~.\eqno(30)$$
Eq.~(30) is the N=2 superfield generalization of the known (and unique) N=1 superfield non-linear 
constraint of ref.~\cite{bg}, describing partial breaking of N=2 extended supersymmetry down to N=1 
supersymmetry in four spacetime dimensions. This correspondence supports the uniqueness of our 
(further extended) action (28) and its 
physical interpretation as the Goldstone action associated with partial breaking of N=4 supersymmetry 
down to N=2 in four spacetime dimensions, with the N=2 vector
multiplet as a Goldstone multiplet ({\it cf.} refs.~\cite{bik,rot}).  

It is also remarkable that our action does {\it not} lead to the propagating 
auxiliary fields $\vec{D}$ despite of the presence of higher derivatives. Though the equations 
of motion for the auxiliary fields may not be algebraic, free kinetic terms for them do not appear, 
with $\vec{D}=0$ being an on-shell solution. Non-vanishing expectation values for 
fermionic or scalar composite operators 
in front of the `dangerous' interacting terms that could lead to a propagation
of the auxiliary fields are also forbidden because of the vanishing N=2 
central charge and unbroken Lorentz- and super-symmetries.

To verify that eq.~(28) is an N=2 supersymmetric extension of the BI action,
it is useful to rewrite it in terms of N=1 superfields by integrating over a 
half of the N=2 superspace anticommuting coordinates. The standard 
identification of the N=1 superspace anticommuting coordinates,~\footnote{
We underline particular values $i=\un{1},\un{2}$ of the internal $SU(2)$ 
indices, and use the N=1 notation \newline ${~~~~~}$ 
$D^2=\frac{1}{2}(D^{\un{1}})^{\a}(D^{\un{1}})_{\a}$ and 
$\bar{D}^2=\frac{1}{2}(\bar{D}_{\un{1}})_{\dt{\a}} 
(\bar{D}_{\un{1}})^{\dt{\a}}$.}
$$ \q^{\a}{}_{\un{1}}=
\q^{\a}~,\quad{\rm and}\quad \bar{\q}_{\dt{\a}}{}^{\un{1}}
=\bar{\q}_{\dt{\a}}~, \eqno(31)$$
implies the N=1 superfield projection rule
$$ G=\left.G(Z)\right|~,\eqno(32)$$
where $|$ means taking a $(\q^{\a}{}\low{\un{2}},
\bar{\q}_{\dt{\a}}{}^{\un{2}})$-independent part
of an N=2 superfield $G(Z)$. As regards the N=2 restricted chiral superfield 
$W$, its N=1 superspace constituents are given by N=1 complex superfields 
$\F$ and $W_{\a}$,
$$ \left.W\right|=\F~, \quad D^{\un{2}}_{\a}\left.W\right|=W_{\a}~,
\quad
-\frac{1}{2}(D^{\un{2}})^{\a}(D^{\un{2}})_{\a}\left.W\right|=
\bar{D}^2\bar{\F}~,\eqno(33)$$
which follow from the N=2 constraints (16). The generalized reality condition 
in eq.~(16) also implies the N=1 superfield Bianchi identity
$$ D^{\a}W_{\a}=\bar{D}_{\dt{\a}}\bar{W}^{\dt{\a}}~,\eqno(34)$$
as well as the relations
$$\eqalign{
 \left.K\right|~=~ & D^2\left( W^{\a}W_{\a} -2\F\bar{D}^2\bar{\F}\right)~.\cr
(\bar{D}_{\un{2}})^{\dt{\a}}\left.K\right|~=~ & 
2iD^2\pa^{\dt{\a}\b}\left(W_{\b}\F\right)~,
\cr
(\bar{D}_{\un{2}})_{\dt{\a}}(\bar{D}_{\un{2}})^{\dt{\a}}\left.K\right|~=~ &
-4D^2\pa_{\m}\left( \F\pa^{\m}\F\right)~,\cr}\eqno(35)$$
together with their conjugates. Eqs.~(33), (34) and (35) are enough to perform
a reduction of any N=2 superspace action depending upon $W$ and $\bar{W}$ into
N=1 superspace by differentiation, 
$$\eqalign{
\int d^4\q ~\to~& \int d^2\q\, \frac{1}{2} (D^{\un{2}})^{\a}(D^{\un{2}})_{\a}~,
\cr
\int d^4\q d^4\bar{\q} ~\to~& \int d^2\q d^2\bar{\q}\, 
\frac{1}{2} (D^{\un{2}})^{\a}
(D^{\un{2}})_{\a}\frac{1}{2} 
(\bar{D}_{\un{2}})_{\dt{\a}}(\bar{D}_{\un{2}})^{\dt{\a}}~.\cr}
\eqno(36)$$
It is now straightforward to calculate the N=1 superfield form of the N=2 
action (28). For our purposes, it is enough to notice that the first term in 
eq.~(28) gives rise to the kinetic terms for the N=1 chiral superfields
 $\F$ and $W_{\a}$,
$$ \int d^2\q\,(-\frac{1}{2}W^{\a}W_{\a} +\F\bar{D}^2\bar{\F})~,\eqno(37)$$
whereas the N=1 vector multiplet contribution arising from the second term in 
eq.~(28) is given by
$$ -\frac{1}{8}\int d^2\q d^2\bar{\q}\,\cy(\left.K\right|,\bar{K}\left.\right|)W^{\a}W_{\a}
\bar{W}_{\dt{\a}}\bar{W}^{\dt{\a}} +\ldots~,\eqno(38)$$
where the dots stand for $\F$-dependent terms. The $W$-dependent contributions
 of eqs.~(37) and (38) exactly coincide with the N=1 supersymmetric extension 
of the BI action found in refs.~\cite{cf,bg} after taking into account that 
the vector field dependence in the first component of the N=1 superfield 
$\left.K\right|$ is given by
$$ \left.K\right| =\left. D^4W^2\right|= 
-\left. 2D^2 ( -\frac{1}{2}W^{\a}W_{\a}+\F\bar{D}^2\bar{\F})\right|
= (F^+)^2-D^2+\ldots, \eqno(39)$$
and similarly for  $\bar{K}\left.\right|$.

The dependence of the N=2 BI action upon the N=1 chiral part $\F$ of the N=2 
vector multiplet is of particular interest, since it is entirely dictated by
N=2 extended supersymmetry and electric-magnetic self-duality. Let's now take 
$W_{\a}=0$ in the N=2 BI action, and  calculate merely the leading terms 
depending upon $\F$ and $\bar{\F}$ in eq.~(28). After some algebra one gets 
the following N=1 superspace action:
$$ S[\F,\bar{\F}]=\int d^4x d^2\q d^2\bar{\q}\left[ \F\bar{\F} 
+4(\F\pa^{\m}\F)(\bar{\F}\pa_{\m}\bar{\F})
-4\pa^{\m}(\F\bar{\F})\pa_{\m}(\F\bar{\F})\right]
+\ldots~,\eqno(40)$$
where the dots stand for the higher order terms depending upon the derivatives
of $\cy$. The field components of the N=1 chiral superfield $\F$ are 
conveniently defined by the projections
$$ \left.\F\right|=\fracmm{1}{\sqrt{2}}\f\equiv\fracmm{1}{\sqrt{2}}(P+iQ)~,
\quad D_{\a}\left.\F\right|=\j_{\a}~,\quad D^2\left.\F\right|=F~,\eqno(41)$$
where $P$ is a real physical scalar, $Q$ is a real physical pseudo-scalar, 
$\j_{\a}$ is a chiral physical spinor, and $F$ is a complex auxiliary field. 
It is not difficult to check that the kinetic terms for the auxiliary 
field components $F$ and $\bar{F}$ cancel in eq.~(40), as they should. This 
allows us to simplify a calculation of the quartic term in eq.~(40) even 
further by going on-shell, i.e. assuming that $\bo\f=F=0$ there, even though 
it is not really necessary. A simple calculation now yields 
$$ S[ \f,\bar{\f} ]=\int d^4x\, \left\{ -\pa^{\m}\f\pa_{\m}\bar{\f} +
2(\pa_{(\m}\f\pa_{\n)}\bar{\f})^2 -(\pa_{\m}\f\pa^{\m}\bar{\f})^2\right\}~,
\eqno(42)$$
which exactly coincides with the leading terms in the derivative expansion of 
the action
$$ S=\int d^4x\left\{1-\sqrt{-\det(\h_{\m\n}+\pa_{\m}P\pa_{\n}P
+\pa_{\m}Q\pa_{\n}Q)}
\right\}~.\eqno(43)$$
This strongly indicates on a possible six-dimensional origin of our 
four-dimensional N=2 supersymmetric BI action, which should be derivable by a 
dimensional reduction from a supersymmetric BI action in six spacetime 
dimensions by identifying the extra two components of a six-dimensional vector
potential with the scalars $P$ and $Q$. The very existence of the super-BI
action in six dimensions is enough to ensure the Goldstone nature of scalars
in eqs.~(42) and (43), as well as the invariance of our N=2 action
with respect to constant shifts of these scalars.

It follows from eq.~(16) that 
$$\bo\left( D^{ij}W-\bar{D}^{ij}\bar{W}\right)=0~,\eqno(44)$$
which implies that the harmonic function ${\rm Im}\,(D^{ij}W)$ is a 
constant,~\footnote{We assume 
that all the superfield components of $W$ are regular in spacetime.} i.e.
$$ D^{ij}W-\bar{D}^{ij}\bar{W}=4iM^{ij}~.\eqno(45)$$
The real constant $\vec{M}$ can be interpreted as a `{\it magnetic\/}' 
Fayet-Iliopoulos term \cite{apt}. This term can be formally removed from the 
constraint (45) by a field redefinition of 
$W$, i.e. at the expense of adding a constant imaginary part to the auxiliary 
scalar triplet $\vec{D}$ of N=2 vector multiplet in eq.~(18). We assume that 
$M=0$ in eq.~(45), which can now be enforced by introducing an unconstrained 
real N=2 superfield Lagrange multiplier 
$\vec{L}=\ha(\vec{\t})\ud{i}{j}L\ud{j}{i}\equiv \ha\tr(\vec{\t}L)$ into the 
action (28) ({\it cf.} ref.~\cite{iz}) as
$$\eqalign{
 S[W] \to S[W,L] ~=~& S[W]+ i\int d^4x d^4\q d^4\bar{\q}\, L_{ij}
\left(D^{ij}W-\bar{D}^{ij}\bar{W}\right)~,\cr
~=~&  S[W] +\left[ i\int d^4x d^4\q\, WW_{\rm magn.} +{\rm h.c.}\right]~,\cr}
\eqno(46)$$
where the N=2 superfield $W$ is a chiral (unrestricted) N=2 superfield, while
$$ W_{\rm magn.}=\bar{D}^4D^{ij}L_{ij} \eqno(47)$$
is the dual or `magnetic' N=2 superfield strength that automatically 
satisfies the N=2 constraints 
(16) due to the defining equation (47). Varying the action (46) with respect 
to $W$, solving the resulting algebraic equation for $W$ in terms of 
$W_{\rm magn.}$ (see e.g., refs.~\cite{bg,rot} for details), and substituting 
the result back into the action (46) results in the dual N=2 action 
$S[W_{\rm magn.}]$ which takes exactly {\it the same} form as that of 
eq.~(28). Hence, it is self-dual with respect to the N=2 supersymmetric 
electric-magnetic duality. Of course, the anticipated supersymmetric BI action
in six spacetime dimensions cannot be self-dual with respect to the 
electric-magnetic duality since the dual to a vector is a vector again in four
spacetime dimensions only.

A massless N=2 (Maxwell) vector multiplet may also be considered as a 
{\it Goldstone} multiplet associated with a partial spontaneous breaking of 
rigid N=4 supersymmetry to N=2 supersymmetry~\cite{bik}. For instance, the
action (28) is obviously invariant under constant shifts of the N=2 superfield 
strength $W$, if $W$ is subject to the on-shell condition $\bo W=0$. Since the 
N=1 supersymmetric BI action \cite{cf} can be interpreted (and, in fact, also 
derived) this way \cite{bg}, it is quite conceivable that a manifestly N=2 
supersymmetric Maxwell-Goldstone action should be equivalent to the N=2 
supersymmetric Nambu-Goto-Born-Infeld action. This implies a hidden invariance
of the action (28) under extra two spontaneously broken and non-linearly 
realized supersymmetries. It is then the full (partially broken) N=4 
supersymmetry that unambiguously determines the entire N=2 supersymmetric BI
action (28) via the non-linear constraint (30). The Goldstone interpretation of
our action (28) is also consistent with the standard interpretation of the 
effective (BPS-like) D-3-brane that breaks a half of supersymmetries in the 
type-IIB superstring theory \cite{aps}. In our case, there are only two
scalars $(P,Q)$ that can be identified with the collective coordinates of
D-3-brane in the directions transverse to its world-volume, so that the
dimension of the ambient spacetime is $1+5$. This case thus corresponds to the
3-brane considered in ref.~\cite{hlp}. 

\section*{Acknowledgements}

I am indebted to J. Bagger for bringing the paper \cite{bg} to my attention,
and E. Ivanov and A. Tseytlin for illuminating discussions and valuable suggestions.
After this work was finished, I was informed by A. Tseytlin that a manifestly N=1 
supersymmetric generalization of the four-dimensional NG action (43) was constructed in 
ref.~\cite{rot}.


\begin{thebibliography}{99}
\bibitem{bi} M. Born and L. Infeld, Proc. Roy. Soc. (London) {\bf A144} (1934)
 425;\\
M. Born, Ann. Inst. Poincar\'e {\bf 7} (1939) 155
\bibitem{open} E. S. Fradkin and A. A. Tseytlin, \pl{163}{85}{123};\\
A. Abouelsaood, C. Callan, C. Nappi and S. Yost, \np{280}{87}{599}
\bibitem{lei} R. G. Leigh, \mpl{4}{89}{2767}
\bibitem{aps} M. Aganagic, C. Popescu and J. H. Schwarz, \np{495}{97}{99} 
\bibitem{eh} W. Heisenberg and H. Euler, Z. Phys. {\bf 98} (1936) 714
\bibitem{schr} E. Schr\"odinger, Proc. Roy. Soc. (London) {\bf A150} (1935) 465
\bibitem{gr} G. W. Gibbons and D. A. Rasheed, \np{454}{95}{185}
\bibitem{tse} A. A. Tseytlin, \np{469}{96}{51}
\bibitem{cf} S. Cecotti and S. Ferrara, \pl{187}{87}{335}
\bibitem{bg} J. Bagger and A. Galperin, \pr{55}{97}{1091}
\bibitem{wess} J. Wess, Acta Physica Austr. {\bf 41} (1975) 409
\bibitem{oldrev} S. V. Ketov, Fortschr. Phys. {\bf 36} (1988) 361
\bibitem{sg} S. J. Gates Jr., and W. Siegel, \np{189}{81}{295}
\bibitem{buch} I. L. Buchbinder, E. I. Buchbinder, E. A. Ivanov, S. M.
Kuzenko and B. A. Ovrut, \pl{412}{97}{309}
\bibitem{krev} S. V. Ketov, {\it Analytic tools to brane technology in N=2
gauge theories mith matter}, DESY and Hannover preprint, hep-th/9806009
\bibitem{bik} S. Bellucci, E. A. Ivanov and S. O. Krivonos, 
{\it Partial breaking N=4 to N=2: hypermultiplet as a Goldstone superfield}, 
Talk given at the 32nd Intern. Symposium on the Theory of Elementary 
Particles, 1--5 September, 1998, Buckow, Germany: to appear in the Proceedings;
hep-th/9809190
\bibitem{rot} M. Ro\v{c}ek and A. A. Tseytlin,
{\it Partial breaking of global D=4 supersymmetry, constrained superfields,
and 3-brane actions}, Imperial College and Stony Brook preprint, November 1998; hep-th/9811232
\bibitem{apt} I. Antoniadis, H. Partouche and T. R. Taylor, \pl{372}{96}{83}
\bibitem{iz} E. A. Ivanov and B. M. Zupnik, {\it Modified N=2 supersymmetry and
Fayet-Iliopoulos terms}, Dubna preprint E2--97--322, November 1997; 
hep-th/9710236
\bibitem{hlp} J. Hughes, J. Liu and J. Polchinski, \pl{180}{86}{370}.

\end{thebibliography}
\end{document}